\documentclass[a4paper,10pt]{article}
\usepackage[utf8x]{inputenc}

\title{Aspects of ABJ Theory }
\author{ Mir Faizal \\
Mathematical Institute, University of Oxford
\\ Oxford
OX1 3LB, United Kingdom 
 }

\begin{document}

\maketitle

\begin{abstract}
In this paper we will analyse the deformation of 
a    ABJ theory in 
harmonic superspace. 
So, we  will first discuss 
deformations of  the harmonic superspace by a graviphoton background and then 
 study the    ABJ theory
in this deformed harmonic superspace. This deformed ABJ theory 
will be shown to posses $\mathcal{N} =6$ supersymmetry. We also discuss
the BRST symmetry of this theory. 
\end{abstract}

\section{Introduction}
 Harmonic 
superspace has been studied in four dimensions \cite{h1, h2}. It has also been 
studied in three dimensions \cite{h21, h3, h4}. In three dimensions it has
$\mathcal{N} = 3$ supersymmetry. Harmonic superspace 
 has been used for analysing the multiple M2-branes  \cite{ahs}. The action for
these multiple M2-branes on $C^4/Z_k$ orbifold is given by 
a theory  called the ABJM theory \cite{1, 2, 3, 4}. 
The ABJM theory has  
 $\mathcal{N} =6$ supersymmetry. However, the
ABJM theory coincides with the  BLG theory 
for the only known example of a Lie $3$-algebra \cite{BL1, BL2, BL3, blG}. 
Thus, the  supersymmetry of the ABJM theory is  expected to be 
enhanced to full $\mathcal{N} =8$ supersymmetry \cite{sabjm, sabjm1}.

The   Chern-Simons 
actions with levels, $k =1,2$ forms the gauge part of  the ABJM theory. 
Furthermore,  the matter fields in the ABJM theory live in the bifundamental representation of the gauge group 
$U(N) \times U(N)$. 
A generalization of the ABJM theory to the  ABJ theory has been made \cite{5, 5ba, 5a1, 5a2, 5a}. In the ABJ theory 
the matter fields live in the 
bifundamental representation of gauge group 
$U(M) \times U(N)$ with $M \neq N$ has been made. 
The ABJ theory also has $\mathcal{N} =6$ supersymmetry.

The presence of $NS$  background causes a noncommutative 
 deformation between the spacetime coordinates   
 \cite{sw, dn, dfr, co}, and the presence of a graviphoton background causes
 a noncommutative deformation between the spacetime and Grassmann coordinates 
 \cite{bgn, gp,  gp01, gp1}. Both these deformations do not break any  
supersymmetry. However, the presence of a $RR$ background causes
 a deformation between the Grassmann  coordinates, and thus,  breaks 
the a certain amount of the supersymmetry
\cite{ov,se,beta1,beta2,beta3,beta4}. 
As M-theory is dual to type II string theory,  a deformation of the string theory 
side will also generate a deformation on the M-theory side. In fact, a 
noncommutative superspace deformation of the ABJM theory in $\mathcal{N} =1$
 superspace have been already studied. 
In this paper we study deformation of the ABJ theory by a graviphoton background
in harmonic superspace. 
  It may be noted that the BRST symmetry of the ABJM theory has been analysed 
in deformed $\mathcal{N} =1$ superspace 
\cite{3, 4, qwer}. 
So, in this paper we will also analyse the 
BRST symmetry of deformed ABJ theory
 in  harmonic superspace. 
\section{Harmonic superspace}
In order to construct harmonic superspace, first 
harmonic variables, $u^{\pm}$, parameterizing the coset $SU(2)/U(1)$ 
are constructed. 
 These harmonic variable are subjected to the 
constraints $
 u^{+i} u^-_i = 1, \,\, \,\, u^{+i} u^+_i = u^{-i} u^-_i =0. 
$
Thus, harmonic  superspace  coordinates are given by 
\begin{equation}
 z = ( x^{ab}, \theta_{a}^{++}, \theta^{--}_a, \theta^0_a, u_i^{\pm} ),
\end{equation}
where $ \theta^{\pm}_a = \theta_a^{ij} u^{\pm}_i u^{\pm}_j$ and $\theta^0_a = \theta^{ij}_a u^+_i u^-_j$. 
In order to construct the harmonic superspace the following derivatives are constructed
\begin{eqnarray}
\nonumber \\ 
{\cal
D}^{++}&=&\partial^{++}+2i\theta^{++ a}\theta^{0 b}
 \partial^A_{ab}
 +\theta^{++a}\frac\partial{\partial\theta^{0 a}}
 +2\theta^{0 a}\frac\partial{\partial\theta^{--a}},\nonumber \\
{  D}^{--}&=&\partial^{--}
 -2i\theta^{--a}\theta^{0 b}\partial^A_{ab}
 +\theta^{--a}\frac\partial{\partial\theta^{0 a}}
 +2\theta^{0 a}\frac\partial{\partial\theta^{++ a}},
\nonumber \\
{  D}^0&=&\partial^0+2\theta^{++ a}\frac\partial{\partial\theta^{++ a}}
-2\theta^{--a}\frac\partial{\partial\theta^{--a}}, 
\end{eqnarray}
and
\begin{eqnarray}
D^{--}_a=\frac\partial{\partial\theta^{++ a}}
 +2i\theta^{--b}\partial^A_{ab}, &&
D^0_a= -\frac12\frac\partial{\partial\theta^{0 a}}
+i\theta^{0 b}\partial^A_{ab},\nonumber \\ 
D^{++}_a=\frac{\partial}{\partial
\theta^{--a}}.&&
\end{eqnarray}
Here $\partial^{++}, \partial^{--}$ and $\partial^{0}$ are given by 
\begin{eqnarray}
&\partial^{++}=u^+_i\frac\partial{\partial u^-_i}, &
\partial^{--}=u^-_i\frac\partial{\partial u^+_i},\nonumber \\ 
&\partial^0 = u^+_i\frac\partial{\partial u^+_i}
-u^-_i\frac\partial{\partial u^-_i}.& 
\end{eqnarray}
Now these derivatives satisfy the following superalbegra, 
\begin{eqnarray}
 \{D^{++}_a, D^{--}_b\}=2i\partial^A_{ab}, \quad \{D^{0}_a,
D^{0}_b\}=-i\partial^A_{ab}, 
\nonumber \\ 
{[{ D}^{\mp\mp}, D^{\pm\pm}_a]}=2D^0_a, \quad [{  D}^{0},
D^{\pm\pm}_a]=\pm 2D^{\pm\pm}_a, 
\nonumber \\ 
\partial^0=[\partial^{++},\partial^{--}],\quad
[{  D}^{++}, {  D}^{--}]={  D}^0. \nonumber \\
\{D^{\pm\pm}_a, D^{0}_b\} = 0\,,\quad [{\cal
D}^{\pm\pm}, D^0_a]=D^{\pm\pm}_a.
  \end{eqnarray}

In harmonic superspace the fields which satisfy, $ D^{++}_a\Phi_A=0 $, are called 
analytic
superfields, $\Phi_A = \Phi_A(\zeta_A)$. Thus, the analytical superfields are  independent of the
$\theta^{--}_a$. 
So, the  coordinates for the analytic subspace 
are given by 
\begin{eqnarray}
\zeta_A=(x^{ab}_A,
\theta^{++}_a, \theta^{0}_a, u^\pm_i), 
  \end{eqnarray}
where
\begin{eqnarray}
x^{ab}_A=(\gamma_m)^{ab}x^m_A=x^{ab}
+i(\theta^{++a}\theta^{--b}+\theta^{++b}\theta^{--a}).
  \end{eqnarray}
The harmonic superspace has $\mathcal{N} =3$ supersymmetry, which is generated by 
\begin{eqnarray}
&Q^{++}_a=u^+_iu^+_j Q_a ^{ij}, &
Q^{--}_a=u^-_iu^-_j Q_a ^{ij},\nonumber \\ 
&Q^0_a = u^+_iu^-_j Q_a ^{ij},& 
\end{eqnarray}
where 
\begin{equation}
 Q_a^{ij} = \frac{\partial}{\partial \theta^a_{ij}} - \theta^{ijb }  \partial_{ab}.
\end{equation}
The measures in the harmonic  superspace are given by  
  \begin{eqnarray}
d^9z &=&-\frac1{16}d^3x
(D^{++})^2 (D^{--})^2(D^{0})^2, \nonumber \\
d\zeta^{(-4)}&=&\frac{1}{4} d^3x_Adu (D^{--})^2(D^{0})^2\,.
  \end{eqnarray}
A conjugation in the  harmonic superspace is defined  by
  \begin{eqnarray}
\widetilde{(u^\pm_i)}=u^{\pm i},\quad \widetilde{(x^m_A)}=x^m_A, \nonumber \\  \widetilde{(\theta^{\pm\pm}_a)}=
\theta^{\pm\pm}_a,\quad \widetilde{(\theta^0_a)}=
\theta^0_a.
  \end{eqnarray}
It is squared to $-1$ on the harmonics  and to $1$ on $x^m_A$  and  
 Grassmann coordinates. So,  the analytic superspace measure is real
 $\widetilde{d\zeta^{(-4)}}=d\zeta^{(-4)}$
and the full superspace measure is imaginary $\widetilde{d^9z}=-d^9z$. 

\section{Deformed ABJ Theory} 
In this section we will construct a   deformed  ABJ theory in the
 harmonic superspace. The deformation will be caused by a graviphoton background. 
We expect a noncommutative deformation  of the 
M-theory to occur in curved backgrounds  due to the three form field 
$C_{\mu\nu\tau}$ and a graviphoton $\psi_{b}^\mu$. If $H_{\mu\nu\tau\rho}$ the field strength of 
this three form field, then  we expect a noncommutative 
deformation proportional to 
 $ (\gamma^\mu\gamma^\nu\gamma^\tau\gamma^\rho H_{\mu\nu\tau\rho})^{ab}\psi_{b}^\mu  $ 
to occur.
We thus deform the harmonic superspace 
 by imposing the following relations, 
\begin{equation}
 \{\hat y_\mu, \hat \theta^{++}_a \} = C^{++}_{\mu a }. \label{p} 
\end{equation}  
Now we start by  
defining a vector field  $V^{++}$ in the harmonic superspace. 
 We can express  this field 
on the deformed superspace as
\begin{eqnarray}
{\hat V^{++}}( \hat z ) 
=
\int dp
\exp i(  p \hat z )  V^{++} (p).
\end{eqnarray}
Now we can similarly express the product of two fields on this deformed superspace. 
We can thus obtain an expression for the   product
  of these fields on ordinary superspace.  This product is given by 
\begin{eqnarray}
V^{++} (z) \star V^{++}  (z) &=&\exp -\frac{1}{2} \left(
C^{++ a\mu}[ \partial^{ ++2}_a \partial^{1}_\mu + \partial^{ ++1}_a \partial^{2}_\mu]\right)
\nonumber\\&& \,\,\,\,\,\,\,\,\,\,\,\times
 {V^{++}}(z_1) { V^{++}}  (z_2)
\left. \right|_{z_1=z_2=z}. 
\end{eqnarray}
It is also  possible to write $V^{--}$ as
\begin{equation}
 V^{--}(z,u)=\sum_{n=1}^\infty (-1)^n \int du_1\ldots
du_n  E^{++},
\end{equation}
where 
\begin{equation}
 E^{++} = \frac{V^{++}(z,u_1)\star V^{++}(z,u_2)\ldots \star 
V^{++}(z,u_n)}{(u^+u^+_1)(u^+_1u^+_2)\ldots (u^+_n u^+)}.
\end{equation}

The action for the deformed ABJ theory is invariant under the gauge group $U(N) \times U(M)$.
Let the the gauge 
fields corresponding to $U(M)$ and $U(N)$ be denoted by  $(V^{++}_L)^A_B$ and
$(V^{++}_R)^{\underline{A}}_{\underline{B}}$, respectively. Thus, indices corresponding to the 
 gauge group $U(M)$ would be denoted  by the  underlined indices. Now the 
 matter fields are given by  $(q^{+})_A^{\underline B}$
 and $(\bar{q}^+)^A_{\underline B}$. 
We can  define  covariant derivatives for
 the matter fields in this deformed theory as
\begin{eqnarray}
\nabla^{++}q^{+}&=&{  D}^{++}q^{+}
 + V^{++}_L \star q^{+}- q^{+} \star V^{++}_R\,,  
\nonumber \\   \nabla^{++}\bar q^{+}&=&{  D}^{++}\bar q^{+}
 -\bar q^{+}  \star V^{++}_L +  V^{++}_R  \star \bar q^{+}\, ,
\end{eqnarray}
The action for the deformed ABJ theory can now be written as 
\begin{equation}
 S = S_{CS, k} [ V^{++}_L]_ \star  + S_{CS, - k} [ V^{++}_R]_ \star   + S_{M} [ q^{+}, \bar q^{+}]_ \star,
\end{equation}
where 
\begin{eqnarray}
 S_{CS, k}[ V^{++}_L]_ \star &=&\frac{ik}{4\pi}\, tr\sum\limits^{\infty}_{n=2} \frac{(-1)^{n}}{n} \int
d^3x d^6\theta du_{1}\ldots du_n  H^{++}_L, \nonumber \\
  S_{CS, -k}[ V^{++}_R]_ \star &=&- \frac{ik}{4\pi}\,tr\sum\limits^{\infty}_{n=2} \frac{(-1)^{n}}{n} \int
d^3x d^6\theta du_{1}\ldots du_n H^{++}_R, \nonumber \\
S_{M} [ q^{+}, \bar q^{+}]_ \star 
&=&tr\int d^3 x d\zeta^{(-4)}
\bar q^{+} \star \nabla^{++}  \star q^{+},
\end{eqnarray}
and
\begin{eqnarray}
 H^{++}_L &=& \frac{V^{++}(z,u_{1} )_L  \star V^{++}(z,u_{2} )_L\ldots
 \star V^{++}(z,u_n )_L }{ (u^+_{1} u^+_{2})\ldots (u^+_n u^+_{1} )},
\nonumber \\ 
H^{++}_R &=& \frac{V^{++}(z,u_{1} )_R  \star V^{++}(z,u_{2} )_R\ldots
 \star V^{++}(z,u_n )_R }{ (u^+_{1} u^+_{2})\ldots (u^+_n u^+_{1} )}.
\end{eqnarray}
The  harmonic superspace used to write the deformed ABJ theory has  manifest $\mathcal{N} =3$
 supersymmetry generated by to the supercharges $Q^{++}_a, Q^{--}_a$ and $Q^{0}_a$. 
However,  the ABJ theory has   $\mathcal{N} =6$ 
supersymmetry.  Thus, we need additional  
additional $\mathcal{N} =3$ supersymmetry.
Thus, this deformed ABJ theory is invariant under the following  supersymmetric transformations, 
\begin{eqnarray}
\delta_\epsilon q^{+}&=& i\epsilon^{a}\hat\nabla^0_a \star q^{+}\,, \nonumber \\
\delta_\epsilon\bar q^{+} &=&i\epsilon^{a} \hat\nabla^0_a \star \bar q^{+ }\,, \nonumber  \\
\delta_\epsilon V^{++}_L&=&\frac{8\pi}k\epsilon^{a}
 \theta^0_a \star q^+\star\bar q^+\,, \nonumber \\
\delta_\epsilon V^{++}_R &=&\frac{8\pi}k\epsilon^{a}
 \theta^0_a \star \bar q^+ \star q^+\,,
\label{epsilon4}
\end{eqnarray}
where
\begin{eqnarray}
 \hat\nabla^0_a \star q^{+}& =& \nabla^0_a \star q^{+} 
+\theta^{--}_a (W^{++}_L \star q^{+} -q^{+} \star W^{++}_R )\,, \nonumber \\
 \nabla^0_a \star q^+&=&D^0_a q^+
 +V^0_{L\, a}\star q^+ -q^+\star V^0_{R\,a }\,, \nonumber \\ V^0_{L\, a}&=&-\frac12D^{++}_a
V^{--}_{L}, \nonumber \\ 
V^0_{R\, a}&=&-\frac12D^{++}_a
V^{--}_{R}.
\end{eqnarray}
The covariant derivatives  $\hat\nabla^0_a \bar q^{+ } $ and $  \nabla^0_a \bar q^{+ }$ are obtained via  conjugation. 
Furthermore, 
the field strengths $W^{++}_R$ and $W^{++}_L$ are given by 
\begin{eqnarray}
 W^{++}_L &=& -\frac{1}{4} D^{++a} D^{++}_{ a}  V^{--}_L, \nonumber \\ 
 W^{++}_R &=& -\frac{1}{4} D^{++a} D^{++}_{ a}  V^{--}_R, 
\end{eqnarray}
where
\begin{eqnarray}
 D^{++} W^{++}_L  + [V^{++}_L , W^{++}_L ]_\star &=0, \nonumber \\ 
 D^{++} W^{++}_R + [V^{++}_R, W^{++}_R]_\star &=0.
\end{eqnarray}
Thus,  we get
\begin{equation}
 \delta_\epsilon  S =0, 
\end{equation}
because   using Fierz rearrangement, we have  $-   \delta_\epsilon 
S_{M} [ q^{+}, \bar q^{+}]_ \star =  \delta_\epsilon S_{CS, k} [ V^{++}_L]_ \star  +  \delta_\epsilon S_{CS, - k} [ V^{++}_R]_ \star $. 
So,  this ABJ theory in the deformed harmonic superspace  has $\mathcal{N} =6$ supersymmetry.

\section{BRST Symmetry}
Now  all the degrees of this deformed ABJ are physical. This is  because 
this theory is invariant under the following  infinitesimal gauge transformations
\begin{eqnarray}
\delta q^{+} &=& \Lambda_L  \star q^{+}-q^{+} \star \Lambda_R,\nonumber \\
 \delta\bar q^{+} &=&\Lambda_R  \star \bar q^{+}-\bar q^{+} \star \Lambda_L,\nonumber \\
\delta V^{++}_L&=&-{  D}^{++}\Lambda_L -[V^{++}_L,\Lambda_L]_ \star,\nonumber \\
\delta V^{++}_R&=&-{  D}^{++}\Lambda_R -[\Lambda_R, V^{++}_R]_ \star.
\end{eqnarray}
As the  deformed ABJ theory is invariant under these gauge transformations, it cannot be quantized  without fixing 
a gauge. We  incorporate the gauge fixing conditions,$
D^{++}   V^{++}_L  =0, \,\, D^{++}  V^{++}_R   =0 
$, at a quantum level by 
 adding gauge fixing and ghost terms  to 
the original action. The gauge fixing term is given by 
\begin{eqnarray}
S_{gf \star} &=& \int d^3 x d\zeta^{(-4)}  tr  \left[b_L \star (D^{++}   V^{++}_L ) + \frac{\alpha}{2}b_L \star b_L \right. \nonumber \\ && \left.  
\,\,\,\,\,\,\,\,\,\,\, -
 b_R  \star (D^{++} V^{++}_R  ) + \frac{\alpha}{2}b _R  \star  b_R 
\right]_|,
\end{eqnarray}
and the ghost term is given by 
\begin{equation}
S_{gh \star} = \int d^3 x d\zeta^{(-4)} tr
[ \overline{c}_L  \star D^{++} \nabla^{++}  \star c_L - \overline{c}_R  \star D^{++} \nabla_a \star c_R ]_|.
\end{equation}
The sum of the original Lagrangian density with the gauge fixing and ghost terms is invariant under the following BRST transformations 
\begin{eqnarray}
s \,V^{++}_L = \nabla^{++} \star  c_L, && s\, V^{++}_R =\nabla^{++}_R  \star   c_R, \nonumber \\
s \,c_L = - {[c_L,c_L]}_ {\star} , && s \,\overline{c}_R =-  b_R - 2 [\overline{c}_R ,  c_R]_{\star}, \nonumber \\
s \, \overline{c}_L = b_L, && s \, c_R = - [c_R ,  c_R]_{\star}, \nonumber \\ 
s \,b_L =0, &&s \, b_R= - [ b_R, \overline{c}_R]_{\star}, \nonumber \\ 
s\, q^{+} = c_L  \star q^{+}-q^{+} \star c_R,&&
s\, \bar q^{+} = c_R  \star \bar q^{+}-\bar q^{+} \star c_L.
\end{eqnarray}
In fact, the sum of the gauge fixing term  and the ghost term,, is  a total BRST variation,
\begin{equation}
S_{gh \star} + S_{gf \star}=   \int d^3 x  d\zeta^{(-4)}   s\, tr  [\Phi]_|
\end{equation}
where 
\begin{equation}
\Phi =  c_L \star \left(D^{++}   V^{++} _L   
 -  \frac{i\alpha}{2}b_L \right) 
-  c_R \star \left(D^{++}   V^{++}_R    
 -  \frac{i\alpha}{2}b_R\right),  
\end{equation}
 As sum  of the  
ghost term and the gauge fixing term  can be expressed 
as a total BRST, it is invariant under the BRST transformations, because 
of the nilpotency of these  transformations, $s^2 =  0$. The 
BRST variation of the original classical Lagrangian density  is its gauge variation with the gauge parameter 
replaced by ghosts. As the original classical Lagrangian density was invariant under the gauge transformations, it 
is also invariant under the BRST transformations. 
So, the effective Lagrangian density which is defined to be a sum of the original classical Lagrangian density, the gauge 
fixing term and the ghost term is invariant under the BRST  transformations. 

As this total Lagrangian  density   
is  invariant under the  BRST   transformations, so 
 we can obtain  
the Norther's charge  $Q$ 
corresponding to the BRST transformations and use 
it to project out the physical states. To do that we first  calculate the 
conserved current corresponding to this symmetry 
\begin{eqnarray}
J^\mu & = & \frac{1}{2}\int d\zeta^{(-4)}  tr 
\left[ \frac{ \partial L_{eff}  }{\partial \mathcal{D}_{\mu} V^{++}_L  } 
\star  s\, 
V^{++}_L  +
 \frac{ \partial L_{eff}  }{\partial  \mathcal{D}_{\mu} c_L } \star  s\, c_L
   +
\frac{ \partial L_{eff}  }{\partial  \mathcal{D}_{\mu} \overline{c}_L }
 \star  s\, \overline{c}_L \right. \nonumber \\&&
 \,\,\,\,\,\, \,\,\,\,\,\,\, \,\,\,\,\,\, \,\,\,\,\,\,\,\,\,\, +
\frac{ \partial L_{eff}  }{\partial \mathcal{D}_{\mu} b_L } \star   s\, b_L
 +
\frac{ \partial L_{eff}  }{\partial \mathcal{D}_{\mu}  V^{++}_R } \star  s\, 
V^{++} _R +
 \frac{ \partial L_{eff}  }{\partial  \mathcal{D}_{\mu}  c_R }
 \star  s\, 
c_R
 \nonumber \\&& 
 \,\,\,\,\,\, \,\,\,\,\,\,\, \,\,\,\,\,\, \,\,\,\,\,\,\,\,\,\, +
\frac{ \partial L_{eff}  }{\partial  \mathcal{D}_{\mu}{\overline{c}}_R }
 \star  s\, \overline{c}_R + 
\frac{ \partial L_{eff}  }{\partial \mathcal{D}_{\mu}  b_R } \star  
 s\,
  b_R + 
\frac{ \partial L_{eff}  }{\partial \mathcal{D}_{\mu}  q^+ } \star  
 s\,
  q^+ 
 \nonumber \\&&\left. 
 \,\,\,\,\,\, \,\,\,\,\,\,\, \,\,\,\,\,\, \,\,\,\,\,\,\,\,\,\, +
\frac{ \partial L_{eff}  }{\partial  \mathcal{D}_{\mu}\bar q^+ }
 \star  s\, \bar q^+ \right]_|,
\end{eqnarray}
where
\begin{equation}
 \int d^3 x d\zeta^{(-4)}  [L_{eff}]_| =  S_{\star} + S_{gh\star}
 + S_{gf \star}.
\end{equation}
Now the conserved BRST charge will be given by 
\begin{equation}
 Q =\int d^3 x \,  J^0.
\end{equation}
As the BRST transformations are nilpotent, 
so for any state $|\phi\rangle$ we have 
\begin{eqnarray}
 Q^2 |\phi\rangle &=& 0.
\end{eqnarray}
The  physical states  $ |\phi_p \rangle $ can now be defined as  
states that are annihilated by $Q$ 
\begin{equation}
 Q |\phi_p \rangle =0. 
\end{equation}
All the states that are obtained by the action of $Q$ on unphysical states are physical. However, they are orthogonal to all physical states. 
Thus, two states differing from each other by addition of such a state will be indistinguishable.
Furthermore, if the asymptotic physical states are given by 
\begin{eqnarray}
 |\phi_{pa,out}\rangle &=& |\phi_{pa}, t \to \infty\rangle, \nonumber \\
 |\phi_{pb,in}\rangle &=& |\phi_{pb}, t \to- \infty\rangle,
\end{eqnarray}
 then a typical $\mathcal{S}$-matrix element can be written as
\begin{equation}
\langle\phi_{pa,out}|\phi_{pb,in}\rangle = \langle\phi_{pa}|\mathcal{S}^{\dagger}\mathcal{S}|\phi_{pb}\rangle.
\end{equation}
Now as the      BRST     
are conserved charges, so they commute with the Hamiltonian 
and thus the time evolution of any physical state will 
also be annihilated by  $Q$, 
\begin{equation}
 Q \mathcal{S} |\phi_{pb}\rangle =0.
\end{equation}  
This implies that the states $\mathcal{S}|\phi_{pb}\rangle$ must be a linear combination of physical states denoted by $|\phi_{p,i}\rangle$.
 So we can write, 
\begin{equation}
\langle\phi_{pa}|\mathcal{S}^{\dagger}\mathcal{S}|\phi_{pb}\rangle
 = \sum_{i}\langle\phi_{pa}|\mathcal{S}^{\dagger}|\phi_{p,i}\rangle
\langle\phi_{p,i}| \mathcal{S}|\phi_{pb}\rangle.
\end{equation}
Since the full $\mathcal{S}$-matrix is unitary this relation implies that the 
 $S$-matrix restricted to
physical sub-space is also unitarity. 

\section{Conclusion}
In this paper we analysed the non-anticommutative deformation of the ABJ theory 
in harmonic superspace. This deformation was caused by a graviphoton backgrounds.  We also 
discuss the BRST symmetry of this theory and used it to show the unitarity of the $S$-matrix. 
There are other type of deformations that can be studied. 
These deformations can occur because of non-vanishing values of anti-commutators between 
Grassmann coordinates and physically correspond to a deformation generated by a $RR$ background
in string theory. As M-theory is dual to string theory, these deformations would also generate 
non-commutative deformations on the M-theory side. 
The interesting thing about these deformations is that they  break some amount of supersymmetry. 
Thus, the some amount of supersymmetry of the 
ABJ theory will be broken by non-anticommutative deformation of the harmonic superspace. 
One such deformation has been recently studied \cite{ncmf}.
 It may be noted that the addition of the mass term 
 breaks 
the superconformal invariance without breaking any supersymmetry
\cite{suprconf}. It will be interesting to analyse these 
aspects further in harmonic superspace.  Furthermore, 
by using a novel Higgsing mechanism, 
 the gauge group of the ABJM theory can
be  spontaneously broken down to its 
diagonal subgroup \cite{zz1,zz2,zz3,zz4a}. Thus, by using this novel Higgsing mechanism the action for 
multiple M2-branes can be reduced to the action for 
multiple
D2-branes.  This analysis has also been performed   in $\mathcal{N} =1$ superspace deformed by 
a graviphoton background \cite{hi}.
It will interesting to analyse this novel Higgsing mechanism in harmonic superspace deformed by a graviphoton 
background.

\end{document}